\documentclass[pra,twocolumn,amsmath,amssymb,floatfi,showpacs]{revtex4}
\usepackage{graphicx}

\begin{document}
\title{\textbf{Dispersion forces inside metallic waveguides}}
\author{Ephraim Shahmoon}
\author{Gershon Kurizki}
\affiliation{Department of Chemical Physics, Weizmann
Institute of Science, Rehovot, 76100, Israel}
\date{\today}

\begin{abstract}
We consider the dispersion energy of a pair of dipoles embedded in a metallic waveguide with transverse dimension $a$ smaller than the characteristic dipolar wavelength. We find that $a$ sets the scale that separates retarded, Casimir-Polder-like, from quasistatic, van der Waals-like, interactions. Whereas in the retarded regime, the energy decays exponentially with inter-dipolar distance, typical of evanescent waves, in the van der Waals regime, the known free-space result is obtained. This short-range scaling implies that the additivity of the dispersion interactions inside a waveguide extends to denser media, along with modifications to related Casimir effects in such structures.
\end{abstract}

\pacs{03.70.+k, 12.20.-m, 42.50.-p, 31.30.J.-} \maketitle

\section{Introduction}
Electric dipoles interact through their electromagnetic fields. Hence, the propagation and scattering properties of these fields in the medium where the dipoles are embedded, strongly influence the spatial dependence of dipole-dipole interactions. In the framework of quantum electrodynamics (QED), these interactions are described as being mediated by virtual photons in different field modes. The space dependence is then determined by the spatial wavefunction of the field modes and hence on the geometry.

When one of the dipoles is initially excited, the resulting interaction induced by virtual photons is  the so-called resonant dipole-dipole interaction (RDDI), where the excitation is dynamically exchanged between the two identical dipoles \cite{LEH,MQED}.
In free space, where the photon modes are plane waves, RDDI scales as $1/r^3$ for inter-dipolar distances $r$, much shorter then the typical dipolar-transition wavelength $\lambda$, and $\cos(2\pi r/\lambda)/r$ for $r\gg \lambda$ \cite{LEH}. Studies of other geometries however, where RDDI is mediated by confined photon modes, e.g. one dimensional surface plasmon polaritons \cite{SPA}, photonic band gaps \cite{KUR},  cavity modes \cite{SEK} and hollow metallic waveguides \cite{LAW,RDDI} to name a few, yielded rather different space-dependencies. For example, in a rectangular hollow metallic waveguide (MWG), depicted in Fig. 1, long range RDDI at distances of $100\lambda$ was found possible, by exploiting the existence of cutoffs in the photon spectrum \cite{RDDI}. For modes with cutoff frequencies higher than $2 \pi c/\lambda$, the interaction is mediated by evanescent modes, i.e. decaying exponents, that sum up to give the free-space result $1/r^3$ at distances much shorter than the size of the waveguide confinement ($a$ and $b$ in Fig. 1) \cite{LAW}.

Consider now the van der Waals (vdW) dispersion interaction, namely when both dipoles are in their ground states, and their mutual interaction is mediated by virtual photons from the vacuum. In free space and for short distances w.r.t the dipole wavelength, $r\ll \lambda$, the vdW interaction scales as $1/r^6$, a direct consequence of the quasistatic limit of the RDDI, $1/r^3$. However, at longer distances, $r\gg \lambda$, in the so-called retarded regime, Casimir and Polder obtained a $1/r^7$ dependence \cite{CP,MQED,MIL}. Different spatial scalings were found however, when two interacting bulk objects rather than point dipoles were considered. Casimir studied the case of two parallel metal plates and found an attraction force that scales as $1/d^4$, with $d$ the distance between the plates \cite{CAS}. Since then, related effects are constantly being studied, mainly considering the interaction between dielectric or metallic objects of various geometries \cite{MIL,REV}. Such experimental \cite{LAM,DEC,MOH,CAP} and theoretical \cite{DAL,EMIG,REY,JOHN,MILT,GRZ1,GRZ2,MAZ} studies
demonstrate the role of retardation and geometry dependence of these vdW-related phenomena \cite{BABB}.

Here however, we would like to take a somewhat different point of view towards the spatial-dependence of the dispersion interaction, reminiscent of that we described for RDDI. Namely, instead of considering objects with different geometries and calculate their interaction energy, we refer back to the original vdW configuration of a pair of point dipoles while changing the geometry of their surrounding environment, and hence the structure of the mediating virtual photon modes. Specifically, here we study dispersion interactions between two dipoles inside a metallic waveguide, in the case where the typical dipole wavelength $\lambda$ is much larger than the typical transverse confinement of the waveguide ($a$ and $b$).

Our results, obtained analytically by QED perturbation theory, are analogous to those of RDDI in a MWG \cite{LAW,RDDI}. They are rather general although full expressions are then derived for the case of a rectangular MWG. Since the smallest cutoff frequency for the mediating transverse modes of the waveguide is of order $c\pi/a$, they become evanescent for the dipole frequency $2 \pi c/\lambda<c\pi/a$. Indeed, we obtain the dispersion energy as a sum of decaying exponentials, $\sum_n b_n e^{-\kappa_n z}$, where $z$ is the inter-dipolar distance on the propagation axis of the MWG, $n$ is an index of a transverse mode, and $b_n,\kappa_n$ are constants. Two distinct regimes are than recognized: the retarded, wave-like, limit, at long distances $z\gg a$, where the interaction decays like $e^{-\kappa_0 z}$, with $\kappa_0$ the smallest of all the $\kappa_n$'s, and the vdW, short-distance limit at $z\ll a$, where the free-space result $1/r^6$ is restored. Hence, as opposed to the free-space case, where $\lambda$ sets the scale for retarded (Casimir-Polder) and non-retarded (vdW) regimes, here we find that the relevant scale is $a$.
Two consequences then emerge: \emph{(1)} unlike free-space, retardation effects are dominant here at distances set by $a$, much shorter than the wavelength $\lambda$, and \emph{(2)} the interaction has a range much shorter than in  free-space due to the exponential decay in the retarded regime. The latter also suggests that the approximate additivity of dipole-dipole and dispersion interactions in a dilute system of many interacting dipoles \cite{MIL}, is expected to hold here at even higher densities.

Another interesting aspect of our results is related to the character of the dominant transverse modes in the Casimir-Polder and vdW regimes. Whereas, the smallest cutoff frequency is obtained for the lowest order transverse-electric (TE) mode and hence it is dominant in the retarded limit, the transverse magnetic (TM) modes are dominant in the short-distance limit and their contributions sum up to give the familiar free-space result.

The paper is organized as follows. In Section II we provide a general discussion and formalism for the calculation of the dispersion energy mediated by photon modes that posses a cutoff, and introduce the modes of the MWG. Next, in Section III, we present the calculation and results for the dispersion energy in a MWG, whereas the comparison with free-space in the corresponding vdW (short-range) and Casimir (retarded) regimes is analyzed in Section IV. Our conclusions are finally given in Section V.
\begin{figure}
\begin{center}
\includegraphics[scale=0.4]{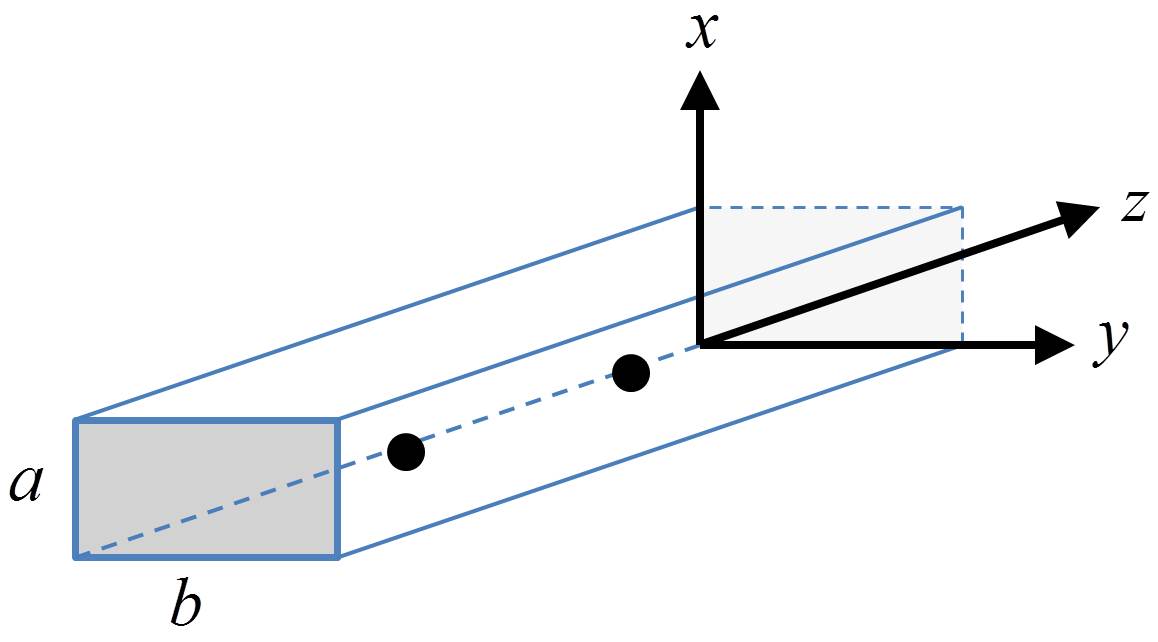}
\caption{\small{(Color online) Dispersion interaction inside a rectangular hollow metallic waveguide. The transverse $x,y$ dimensions of the waveguide are $a$ and $b$, respectively. A pair of interacting dipoles, represented by black dots, are located inside the waveguide and along its propagation axis $z$.
}}
\end{center}
\end{figure}

\section{Dispersion interaction via modes with cutoffs}
\subsection{The system}
We consider two identical dipoles, e.g. made of atoms or molecules, with ground state $|g\rangle$ and a ladder of excited levels $\{|e\rangle\}$ with corresponding energies $E_e$ w.r.t that of the ground state. The excited levels possess dipole-transition matrix elements $\mathbf{d}_e$ to the ground state. The dipoles are placed inside a hollow waveguide that supports photon modes with normalized mode functions,
\begin{equation}
\mathbf{u}^{\mu}_{mn k}(\mathbf{r})=\frac{1}{\sqrt{L}}e^{i k z} \mathbf{E}^{\mu}_{mn k}(x,y).
\label{u}
\end{equation}
Here $k$ is the wavenumber at the propagation axis $z$ with quantization length $L$, $\mu$ is the index of the polarization and $m$ and $n$ are indices of the transverse mode whose dependence on the transverse coordinates is described by $\mathbf{E}^{\mu}_{mn k}(x,y)$. Each of the transverse modes $\mu_{mn}$ has a dispersion relation
\begin{equation}
\omega^{\mu}_{mnk}=c\sqrt{k_{mn}^2+k^2},
\label{w}
\end{equation}
where $c$ is the speed of light in the material that fills the hollow waveguide and $ck_{mn}$ is the cutoff frequency of this mode. The Hamiltonian of the two dipoles $1,2$ and the field modes is then described by
\begin{eqnarray}
H&=&H_0+H_I,
\nonumber \\
H_0&=&\sum_{\nu=1,2}\sum_{e_{\nu}}E_{e_{\nu}} |e_{\nu}\rangle \langle e_{\nu}|+\sum_{\mu m n k}\hbar\omega^{\mu}_{mnk}\hat{a}^{\mu \dag}_{mnk}\hat{a}^{\mu}_{mnk},
\nonumber\\
H_I&=&-\hbar\sum_{\nu=1}^2\sum_{e_{\nu}}\sum_{\mu mnk}\left(|e_{\nu}\rangle \langle g|+\mathrm{h.c.}\right)\left(i g^{ m n k}_{\mu, \nu e} \hat{a}^{\mu }_{mnk} +\mathrm{h.c.}\right),
\nonumber\\
\label{H}
\end{eqnarray}
where $g^{ m n k}_{\mu, \nu e}=\sqrt{\frac{\omega^{\mu}_{mnk}}{2\epsilon \hbar}}\mathbf{d}_{e_{\nu}}\cdot\mathbf{u}^{\mu}_{mnk}(\mathbf{r}_{\nu})$ is the dipole coupling between the $|e\rangle\leftrightarrow|g\rangle$ transition of atom $\nu=1,2$ located at $\mathbf{r}_{\nu}$, and the $\mu_{mn}k$ mode with lowering operator $ \hat{a}^{\mu }_{mnk}$, and $\epsilon$ being the permittivity inside the MWG.

\subsection{QED perturbation theory}
In order to calculate the dispersion interaction energy we follow the QED perturbative approach of Ref. \cite{MQED}. The interaction energy $U$ is obtained as the fourth order correction to the energy of the ground state of the system, $|G\rangle=|g_1,g_2,0\rangle$ with energy $E_G=0$, where both atoms are in their ground state and the photon modes are in the vacuum $|0\rangle$,
\begin{equation}
U=-\sum_{I_1,I_2,I_3}\frac{\langle G|H_I|I_3\rangle \langle I_3|H_I|I_2\rangle \langle I_2|H_I|I_1\rangle \langle I_1|H_I|G\rangle}{(E_{I_1}-E_G)(E_{I_2}-E_G)(E_{I_3}-E_G)}.
\label{E4}
\end{equation}
Here $|I_j\rangle$ are intermediate (virtual) states, and $E_q$ is the free Hamiltonian ($H_0$) energy of the state $|q\rangle$.
The above sum contains 12 possible terms, each of which describes a different virtual process and can be represented by a diagram. In Fig. 2 we present the diagrams of two of these processes: the one in (a) includes the intermediate states $|I_1\rangle=|e_1,g_2,1^{\mu}_{mnk}\rangle ,\, |I_2\rangle=|e_1,e_2,0\rangle$ and $|I_3\rangle=|g_1,e_2,1^{\mu'}_{m'n'k'}\rangle$, with $|1^{\mu}_{mnk}\rangle=\hat{a}^{\mu \dag}_{mnk}|0\rangle$, whereas the one in (b) describes the process $|I_1\rangle=|e_1,g_2,1^{\mu}_{mnk}\rangle ,\, |I_2\rangle=|g_1,g_2,1^{\mu}_{mnk}1^{\mu'}_{m'n'k'}\rangle$ and $|I_3\rangle=|g_1,e_2,1^{\mu'}_{m'n'k'}\rangle$.
\begin{figure}
\begin{center}
\includegraphics[scale=0.43]{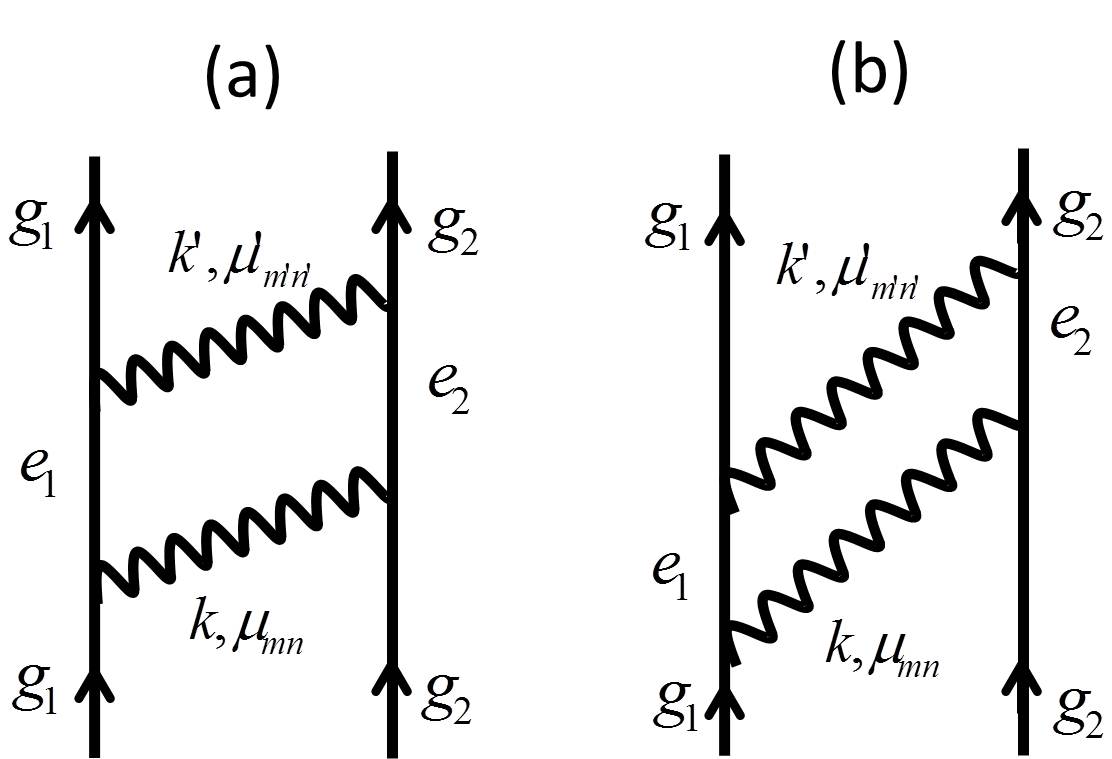}
\caption{\small{(Color online) Dispersion energy calculation by QED perturbation theory. Two of 12 possible processes that contribute to the energy correction of the state $|G\rangle=|g_1,g_2,0\rangle$, Eq. (\ref{E4}). (a) Diagram of the process that includes the intermediate states $|I_1\rangle=|e_1,g_2,1^{\mu}_{mnk}\rangle ,\, |I_2\rangle=|e_1,e_2,0\rangle$ and $|I_3\rangle=|g_1,e_2,1^{\mu'}_{m'n'k'}\rangle$, with $|1^{\mu}_{mnk}\rangle=\hat{a}^{\mu \dag}_{mnk}|0\rangle$. Its corresponding term in the sum Eq. (\ref{E4}) has the denominator $D_a$ from Eq. (\ref{D}).
(b) Here the intermediate states are $|I_1\rangle=|e_1,g_2,1^{\mu}_{mnk}\rangle ,\, |I_2\rangle=|g_1,g_2,1^{\mu}_{mnk}1^{\mu'}_{m'n'k'}\rangle$ and $|I_3\rangle=|g_1,e_2,1^{\mu'}_{m'n'k'}\rangle$, with the denominator $D_b$ from (\ref{D}).
}} \label{QED}
\end{center}
\end{figure}
It is easy to show that all 12 terms in the sum (\ref{E4}) have the same numerator, but are distinct in  their denominators. For example, the terms that correspond to the diagrams in Figs. 2(a) and 2(b) possess the denominators,
\begin{eqnarray}
D_a&=&(\hbar\omega^{\mu}_{mnk}+E_{e_1})(E_{e_1}+E_{e_2})(\hbar\omega^{\mu'}_{m'n'k'}+E_{e_2}),
\nonumber\\
D_b&=&(\hbar\omega^{\mu}_{mnk}+E_{e_1})(\hbar\omega^{\mu}_{mnk}+\hbar\omega^{\mu'}_{m'n'k'})(\hbar\omega^{\mu'}_{m'n'k'}+E_{e_2}),
\nonumber\\
\label{D}
\end{eqnarray}
respectively.
\subsubsection{The case of tight confinement: $a,b\ll 2\pi\hbar c/E_e$}
Let us now consider the case where the wavelengths associated with all the dipolar transitions, $\lambda_e=2\pi\hbar c/E_e$ are much larger than the transverse confinement of the waveguide, $a$ and $b$. Since the cutoff wavenumbers of the transverse modes, $k_{mn}$, result from the confinement, they must satisfy $k_{mn}\geq \pi/a$ or $\pi/b$ $\forall \mu_{mn}$, and we obtain in our case $k_{mn}\gg E_e/(\hbar c)$ for all states $|e\rangle$ and modes $\mu_{mn}$. Then, considering also the dispersion relation, Eq. (\ref{w}), we find
\begin{equation}
\omega^{\mu}_{mnk}\gg E_e/\hbar \quad \forall e,\mu_{mn},k.
\label{con}
\end{equation}
This means that $D_a\ll D_b$ such that the contribution of diagram (b) is negligible w.r.t that of diagram (a) from Fig. 2. In fact, there are three more diagrams with denominator $D_a$ and in a similar fashion one can show that $D_a$ is smaller than the rest of the possible denominators of the remaining 7 diagrams, out of the total 12.  Hence, the dispersion energy in the tight-confinement case can be approximated by considering only the contribution of the  diagrams with $D_a$. Namely, we obtain the energy $U$ by 4 times the contribution of diagram (a),
\begin{eqnarray}
U&\approx&-\frac{1}{(2\pi\epsilon)^2}\sum_{e_1,e_2}\sum_{ijlq} \frac{d_{e_2 i}d_{e_1 j}d_{e_2 l}d_{e_1 q}}{E_{e_1}+E_{e_2}}F_{e_2 ij}F_{e_1 lq},
\nonumber\\
F_{eij}&=&\sum_{\mu mn}F^{\mu}_{mn,eij},
\nonumber\\
F^{\mu}_{mn,eij}&=&\int_{-\infty}^{\infty}dk\frac{\omega^{\mu}_{mnk}}{\omega^{\mu}_{mnk}+E_e/\hbar}
\nonumber\\
&&\times E^{\mu}_{mnk,i}(x_2,y_2)E^{\mu\ast}_{mnk,j}(x_1,y_1)e^{ikz}.
\label{U}
\end{eqnarray}
Here $z$ is the inter-dipolar distance in the waveguide propagation axis $z$, whereas ($x_{\nu},y_{\nu}$) is the transverse position of dipole $\nu$. The indices $i,j,l,q$ run over the projections of the vectors $\mathbf{d}_e$ and $\mathbf{E}^{\mu}_{mnk}(x,y)$ onto the $x,y,z$ directions, e.g. $d_{e x}=\mathbf{d}_e\cdot\mathbf{e}_x$. Further simplification can be made by applying the inequality (\ref{con}) in the above expression for $F^{\mu}_{mn,eij}$, obtaining
\begin{equation}
F^{\mu}_{mn,eij}\approx\int_{-\infty}^{\infty}dk E^{\mu}_{mnk,i}(x_2,y_2)E^{\mu\ast}_{mnk,j}(x_1,y_1)e^{ikz}.
\label{F}
\end{equation}
Eq. (\ref{U}) shows that the spatial dependence of the dispersion energy $U$ is encoded in $F^{\mu}_{mn,eij}$, while Eq. (\ref{F}) further simplifies its calculation by the Fourier transform of a multiplication of the waveguide modes' transverse profiles.

\subsection{The modes of a MWG}
We consider the rectangular hollow MWG shown in Fig. 1, whereas any of our calculations can be easily generalized to e.g. a circular hollow MWG. We also take the limit of a perfect metal, which is reasonable when the relevant frequencies $E_e/\hbar$, $c\pi/a$ and $c\pi/b$ are small enough w.r.t the plasma frequency of the metal. There are two possible polarizations for the transverse modes, TE and TM \cite{KONG}, and by normalizing their transverse profiles, i.e. demanding $\int_0^a dx\int_0^b dy  \mathbf{E}^{\mu \ast}_{mnk}(x,y)\cdot \mathbf{E}^{\mu}_{mnk}(x,y)=1$, we obtain
\begin{eqnarray}
&&\mathbf{E}^{TM}_{mnk}(x,y)=\frac{2c}{\sqrt{A}}\left(\frac{k_{mn}}{\omega^{TM}_{mnk}}\sin\left(\frac{m\pi}{a}x\right)\sin\left(\frac{n\pi}{b}y\right)\mathbf{e}_z
\right. \nonumber \\ &&\left.
+\frac{i k}{k_{mn} \omega^{TM}_{mnk}}\left[\frac{\pi}{a}m\cos\left(\frac{m\pi}{a}x\right)\sin\left(\frac{n\pi}{b}y\right)\mathbf{e}_x
\right. \right. \nonumber \\ &&\left. \left.
+\frac{\pi}{b} n \sin\left(\frac{m\pi}{a}x\right)\cos\left(\frac{n\pi}{b}y\right)\mathbf{e}_y \right]\right),
\nonumber \\
&&\mathbf{E}^{TE}_{mn}(x,y)=\frac{2}{\sqrt{A}k_{mn}} \left[-\frac{\pi}{b}n\cos\left(\frac{m\pi}{a}x\right)\sin\left(\frac{n\pi}{b}y\right)\mathbf{e}_x
\right. \nonumber \\ &&\left.
    +\frac{\pi}{a} m \sin\left(\frac{m\pi}{a}x\right)\cos\left(\frac{n\pi}{b}y\right)\mathbf{e}_y \right],
\nonumber \\
\label{Emn}
\end{eqnarray}
where $A=ab$ is the transverse area of the waveguide, and we note that the transverse profile of the TE modes, $\mathbf{E}^{TE}_{mn}$, does not depend on the wavenumber $k$ (or on frequency), hence the index $k$ is omitted.
The mode frequencies $\omega^{\mu}_{mnk}$ with $\mu=TM,TE$ are those from Eq. (\ref{w}) with the  cutoff wavenumbers $k_{mn}$ given by
\begin{equation}
k_{mn}=\sqrt{(m\pi/a)^2+(n\pi/b)^2}.
\label{kmn}
\end{equation}
The indices $m,n$ are positive integers, whereas the lowest modes are $TM_{11}$ and $TE_{01}$ or $TE_{10}$ for TM and TE, respectively.

\section{energy calculation and general results}
In this section we present the calculation and general results for $F^{\mu}_{mn,eij}$ form Eqs. (\ref{U}) and (\ref{F}), for all dipole orientations $i,j$ and mode polarizations $\mu=TM,TE$.
\subsection{TM modes}
We begin with the $i,j=z,z$ case, namely insert the $z$ component of the TM mode $E^{TM}_{mnk,z}(x,y)=\mathbf{E}^{TM}_{mnk}(x,y)\cdot \mathbf{e}_z$ from Eq. (\ref{Emn}) into the integral in Eq. (\ref{F}). We get,
\begin{eqnarray}
F^{TM}_{mn,ezz}&=&\frac{4}{A}k_{mn}\sin^2\left(\frac{m\pi}{a}x\right)\sin^2\left(\frac{n\pi}{b}y\right)
\nonumber \\
&&\times\int_{-\infty}^{\infty} du\frac{1}{u^2+1}e^{i u k_{mn} z},
\label{TMzz1}
\end{eqnarray}
where $u$ is a dimensionless integration variable. Here we assumed for simplicity $(x_1,y_1)=(x_2,y_2)\equiv(x,y)$.
The above integral is simply the Fourier transform of a Lorentzian, giving rise to an exponential decay with distance $z$,
\begin{equation}
F^{TM}_{mn,ezz}=\frac{4\pi}{A}k_{mn}\sin^2\left(\frac{m\pi}{a}x\right)\sin^2\left(\frac{n\pi}{b}y\right)e^{-k_{mn} z}.
\label{TMzz}
\end{equation}
The typical range of the $TM_{mn}$-mediated interaction between the $z$ component of the dipoles is than set by $k_{mn}^{-1}$, which cannot exceed $k_{11}^{-1}=\sqrt{a^2+b^2}/\pi$. The exponential decay is a consequence of interaction via evanescent modes, since all of the frequencies of the dipole transitions $E_e/\hbar$, were assumed here to be lower than all the cutoff frequencies of the MWG, $ck_{mn}$ (see Sec. IIB).

Turning to the  $i,j=x,x$ case, we insert $E^{TM}_{mnk,x}(x,y)$ from Eq. (\ref{Emn}) into Eq. (\ref{F}) and obtain,
\begin{eqnarray}
F^{TM}_{mn,exx}&=&\frac{4}{A}k_{mn}\left(\frac{m\pi}{k_{mn}a}\right)^2\cos^2\left(\frac{m\pi}{a}x\right)\sin^2\left(\frac{n\pi}{b}y\right)
\nonumber \\
&&\times\int_{-\infty}^{\infty} du\frac{u^2}{u^2+1}e^{i u k_{mn} z}.
\label{TMxx1}
\end{eqnarray}
We perform the integral by contour integration with regularization and get,
\begin{equation}
F^{TM}_{mn,exx}=\frac{4\pi}{A}k_{mn}\left(\frac{m\pi}{k_{mn}a}\right)^2\cos^2\left(\frac{m\pi}{a}x\right)\sin^2\left(\frac{n\pi}{b}y\right)e^{-k_{mn} z},
\label{TMxx}
\end{equation}
obtaining again exponential decay with the exponent $k_{mn}$. We proceed in a similar fashion for the rest of the options and find
\begin{eqnarray}
F^{TM}_{mn,eyy}&=&\frac{4\pi}{A}k_{mn}\left(\frac{n\pi}{k_{mn}b}\right)^2\sin^2\left(\frac{m\pi}{a}x\right)\cos^2\left(\frac{n\pi}{b}y\right)e^{-k_{mn} z},
\nonumber\\
F^{TM}_{mn,exy}&=&-\frac{\pi^2}{2A^2k_{mn}}\sin\left(\frac{2m\pi}{a}x\right)\sin\left(\frac{2n\pi}{b}y\right))e^{-k_{mn} z},
\nonumber \\
F^{TM}_{mn,exz}&=&-\frac{2\pi^2}{A a}m\sin\left(\frac{2m\pi}{a}x\right)\sin^2\left(\frac{n\pi}{b}y\right))e^{-k_{mn} z},
\nonumber\\
F^{TM}_{mn,eyz}&=&-\frac{2\pi^2}{A b}n\sin^2\left(\frac{m\pi}{a}x\right)\sin\left(\frac{2n\pi}{b}y\right))e^{-k_{mn} z},
\label{TM}
\end{eqnarray}
and $F^{TM}_{mn,eij}=F^{TM}_{mn,eji}$ for $(x_1,y_1)=(x_2,y_2)\equiv(x,y)$.

\subsection{TE modes}
By virtue of Eq. (\ref{Emn}) we note that $E^{TE}_{mn,i}(x,y)$ ($i=x,y$) is independent of $k$ , such that it can be taken out of the integral in Eq. (\ref{U}) for $F^{TE}_{mn,eij}$. Then, by using Eq. (\ref{w}), we are left to solve the integral,
\begin{equation}
I=\int_{-\infty}^{\infty} du\frac{\sqrt{u^2+1}}{\sqrt{u^2+1}+u_e}e^{i \zeta u},
\label{I1}
\end{equation}
with $\zeta=k_{mn} z$ and $u_e=E_e/(\hbar c k_{mn})$. By contour integration methods, this integral is found to be equal to
\begin{eqnarray}
I&=&-2u_e \int_{1}^{\infty} du\frac{\sqrt{u^2-1}}{u^2-1+u_e^2}e^{- \zeta u}
\nonumber\\
&&\approx -2u_e \int_{1}^{\infty} du\frac{1}{\sqrt{u^2-1}}e^{-\zeta u}=-2u_eK_0(\zeta),
\label{I2}
\end{eqnarray}
where $K_0(u)$ is the zeroth order modified Bessel function and where the lowest order approximation due to the tight confinement, $u_e\ll 1$, was taken. We thus obtain for any $i=x,y$ and $j=x,y$,
\begin{equation}
F^{TE}_{mn,eij}=E^{TE}_{mn,i}(x_2,y_2)E^{TE}_{mn,j}(x_1,y_1)\frac{E_e}{\hbar c} K_0(k_{mn} z).
\label{TE}
\end{equation}

\section{Comparison to free-space: vdW and Casimir-Polder regimes}
After we obtained the general expression for the dispersion energy for every polarization of the dipoles and mediated by any of the transverse modes $\mu_{mn}$, we now wish to consider two relevant limits; i.e., the retarded, Casimir-Polder limit, and the quasistatic, vdW limit, and compare the results with their free-space counterparts.

\subsection{Casimir-Polder: $z\gg a,b$}
In the tight confinement case we consider, the scale for the on-axis distance $z$ at which wave effects emerge is set by the confinement $a,b$. However, since $a,b\ll \lambda_e$, wave effects in the confined case mean decay rather than oscillation with $z$ due to the evanescent nature of the modes.

\subsubsection{TM contribution}
Eqs. (\ref{TMzz}),(\ref{TMxx}) and (\ref{TM}) show that for any dipole orientation, the TM-mediated energy decays exponentially with the exponent $k_{mn}$ like,
\begin{equation}
\sim\exp(-\pi\frac{z}{a}\sqrt{m^2+n^2}),
\label{TMCP1}
\end{equation}
where $b\sim a$ is assumed here for simplicity. Clearly, when $z\gg a$, the contribution of the lowest order TM mode, namely $m=1$ and $n=1$, overwhelms that of the rest of the TM modes, and the $\mu=TM$ contribution to the sum for $F_{eij}$ in Eq. (\ref{U}), can be taken solely as $F^{TM}_{11,eij}$, where
\begin{equation}
F^{TM}_{11,eij}\propto e^{-\sqrt{2}\pi z/a}.
\label{TMCP}
\end{equation}

\subsubsection{TE contribution}
The argument of the modified Bessel function in Eq. (\ref{TE}), is exactly that of the exponent in (\ref{TMCP1}), and in the retarded regime $z\gg a$, it is clearly much larger than 1 for any $m,n\neq 0,0$ (as in TE). We then use the approximation $K_0(k_{mn}z)\approx \sqrt{\pi/(2k_{mn} z)}e^{-k_{mn}z}$ for $k_{mn} z\gg 1$. The contribution of the TE modes can then be approximated by that of the two dominant lowest order modes; i.e. $TE_{01}$ and $TE_{10}$, where
\begin{equation}
F^{TE}_{10,eij}\propto  \frac{e^{-\pi z/a}}{\sqrt{\pi z/a}},
\quad
F^{TE}_{01,eij}\propto  \frac{e^{-\pi z/b}}{\sqrt{\pi z/b}}.
\label{TECP}
\end{equation}

\subsubsection{Total dispersion energy}
From Eqs. (\ref{TMCP}) and  (\ref{TECP}) we conclude that the TE contribution is more dominant than the TM one due to its slower decay with $z/a$. The total energy $U$ from (\ref{U}), in the retarded limit, can thus be approximated by that of the $TE_{01}$ and $TE_{10}$ modes,
\begin{eqnarray}
U&\approx&-\frac{1}{(2\pi\epsilon)^2}\sum_{e_1,e_2}\sum_{ijlq} \frac{d_{e_2 i}d_{e_1 j}d_{e_2 l}d_{e_1 q}}{E_{e_1}+E_{e_2}} \left[F^{TE}_{10,e_2ij}F^{TE}_{10,e_1lq}
\right. \nonumber \\ &&\left.
+F^{TE}_{10,e_2ij}F^{TE}_{01,e_1lq}+F^{TE}_{01,e_2ij}F^{TE}_{10,e_1lq}+F^{TE}_{01,e_2ij}F^{TE}_{01e_1lq}\right].
\nonumber\\
\label{UCP1}
\end{eqnarray}
Using Eq. (\ref{Emn}), we find
\begin{equation}
E^{TE}_{01,i}=-\frac{2}{\sqrt{A}}\sin(\frac{\pi}{b}y)\delta_{ix},
\quad
E^{TE}_{10,i}=-\frac{2}{\sqrt{A}}\sin(\frac{\pi}{a}x)\delta_{iy},
\label{TE01}
\end{equation}
where $\delta_{ij}$ is the Kronecker delta. In order to illustrate our results, let us take $x_1=x_2=y_1=y_2\equiv x$ and $a=b$, and assume the dipoles are randomly oriented in space in a uniform distribution, i.e. $\langle d_{ex}^2 \rangle=\langle d_{ey}^2 \rangle=\langle d_{ez}^2 \rangle=(1/3) |\mathbf{d}_e|^2 $. We then obtain
\begin{eqnarray}
U&\approx&-\frac{8\pi^2}{9}\sin^4\left(\frac{\pi}{a}x\right)\frac{1}{\epsilon^2}\sum_{e_1,e_2} \frac{|\mathbf{d}_{e_1}|^2|\mathbf{d}_{e_2}|^2}{E_{e_1}+E_{e_2}}
\nonumber \\
&& \times \frac{1}{\lambda_{e_1}\lambda_{e_2}a^3}\frac{1}{z}e^{-2\pi z/a},
\nonumber\\
\label{UCP}
\end{eqnarray}
where $\lambda_e= 2\pi\hbar c/E_e$ is as usual the wavelength associated with the $|g\rangle\leftrightarrow |e\rangle$ transition. We note that the effect of the transverse position of both dipoles, $x$, is contained in the numerical prefactor $\sin^4\left(\frac{\pi}{a}x\right)$ and becomes equal to $1$ when the dipoles are in the center, $x=a/2$. Using the atomic dynamic polarizability  $\alpha(\omega)=(2/3)\sum_e E_e |\mathbf{d}_{e}|^2/[E_e^2-(\hbar\omega)^2]$, the dispersion energy (\ref{UCP}) can be written as \cite{MQED}
\begin{eqnarray}
U&\approx&-2\pi\sin^4\left(\frac{\pi}{a}x\right)\frac{1}{\epsilon^2}\int_{-\infty}^{\infty} du \alpha_1(iu) \alpha_2(iu)
\nonumber \\
&& \times \frac{1}{\lambda_{e_1}\lambda_{e_2}a^3}\frac{1}{z}e^{-2\pi z/a},
\nonumber\\
\label{UCP2}
\end{eqnarray}
with $u=i\hbar\omega$.

\subsubsection{Comparison to free-space result}
It is now instructive to compare the above result to its free-space counterpart, $U_{fs}$, in two different regimes. For $a\ll z \ll \lambda_e$ $\forall e$, the corresponding result in free-space is that of the non-retarded, vdW, limit \cite{MQED}
\begin{equation}
U_{fs}\approx-\frac{1}{24\pi^2}\frac{1}{\epsilon^2} \sum_{e_1,e_2} \frac{|\mathbf{d}_{e_1}|^2|\mathbf{d}_{e_2}|^2}{E_{e_1}+E_{e_2}} \frac{1}{r^6},
\label{Ufs1}
\end{equation}
where $r$, the inter-dipolar distance, is equal to $z$ in our case. For a single excited level $e_1=e_2=e$ and $x=a/2$, the ratio between the MWG and free-space results becomes,
\begin{equation}
\left.\frac{U}{U_{fs}}\right|_{e_{1,2}=e}\approx \frac{64\pi^4}{3}\frac{z^5}{\lambda_e^2a^3}e^{-2\pi z/a}.
\label{R1}
\end{equation}
While the factor $z^5/(\lambda_e^2a^3)$ may become larger than $1$, since $z$ is between $a$ and $\lambda_e$, the exponential decay still makes this ratio much smaller than $1$, such that the free-space energy is much stronger than the one mediated by the MWG in this $z$ limit. This is plotted in Fig. 3(a) for $\lambda_e/a=100$.

Next, we consider the limit $z\gg \lambda_e$ $\forall e$, where the free-space result takes the retarded, Casimir-Polder form \cite{MQED}
\begin{equation}
U_{fs}\approx\frac{23}{144\pi^3}\frac{\hbar c}{\epsilon^2} \sum_{e_1,e_2} \frac{|\mathbf{d}_{e_1}|^2|\mathbf{d}_{e_2}|^2}{E_{e_1}E_{e_2}} \frac{1}{r^7}.
\label{Ufs2}
\end{equation}
Taking again $r=z$, $e_1=e_2=e$ and $x=a/2$, the MWG to free space ratio becomes,
\begin{equation}
\left.\frac{U}{U_{fs}}\right|_{e_{1,2}=e}\approx \frac{128\pi^6}{23}\frac{z^6}{\lambda_e^3a^3}e^{-2\pi z/a}.
\label{R2}
\end{equation}
Although $z^6/(\lambda_e^3a^3)\gg 1$, it is again the exponential decay that makes this ratio go practically to zero, as can be seen on Fig 3(b), for $\lambda_e/a=10$.
\begin{figure}
\begin{center}
\includegraphics[scale=0.3]{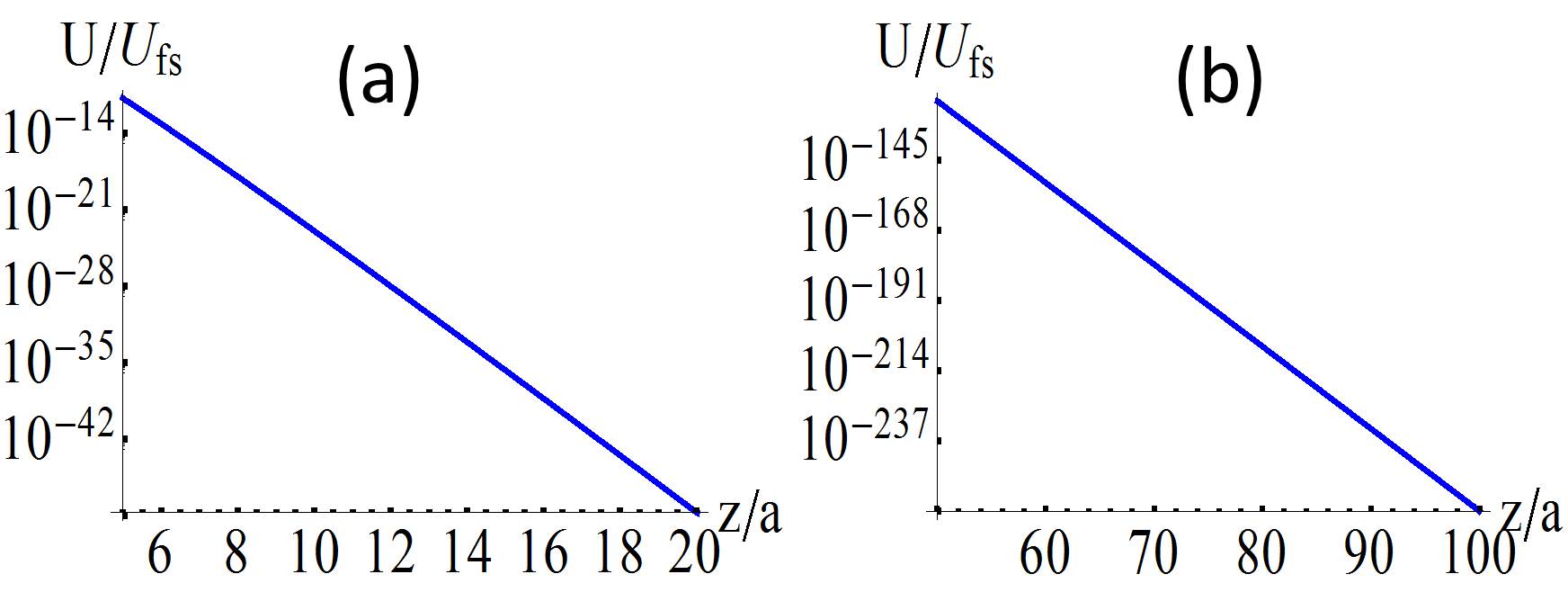}
\caption{\small{(Color online) Dispersion energy in the metal waveguide (MWG) $U$ compared to its free-space counterpart $U_{fs}$, in the retarded $z\gg a$ regime. Here a single dipole transition with wavelength $\lambda_e$ is assumed. (a) $a\ll z \ll \lambda_e$: in this limit the free-space energy is in the vdW regime, Eq. (\ref{Ufs1}), and the ratio $U/U_{fs}$ from Eq. (\ref{R1}) is plotted for $\lambda_e=100a$. (b) $z\gg \lambda_e$: in this limit the free-space energy takes the Casimir-Polder form [Eq. (\ref{Ufs2})] and the ratio $U/U_{fs}$ from Eq. (\ref{R2}) is plotted for $\lambda_e=10a$. Both plots are linear in a semi-log scale, hence they show that the free-space energy is exponentially and many orders of magnitude larger than the short-range interaction mediated by the MWG modes, as suggested by Eqs. (\ref{R1}) and (\ref{R2}).
}}
\end{center}
\end{figure}

Eqs. (\ref{R1}) and (\ref{R2}) imply that the dispersion interaction in a MWG is much shorter ranged than that in free space. This is further discussed in the conclusions in section V.

\subsection{vdW: $z\ll a,b$}
We define the vdW regime as the range of inter-dipolar distance $z$ where wave effects are not significant. This is the case for $z\ll a,b$, where the dipoles are close enough such that the exponential decay of the TE and TM evanescent waves, set by $a$ and $b$, is almost not felt. Moreover, when the dipoles are placed in the center of the waveguide and are close enough such that they do not "sense" its metallic plates, free-space behavior is expected to be observed.

\subsubsection{TM contribution}
The $TM_{mn}$ terms [Eqs. (\ref{TMzz}),(\ref{TMxx}) and (\ref{TM})] possess an exponential decay $e^{-\sqrt{m^2+n^2}\pi z/a}$, which in the $z\ll a,b$ is slowly decaying with $m$ and $n$. Hence, unlike the retarded regime, where only the $TM_{11}$ was considered, here we must consider all terms if we wish to account for arbitrary small $z$. In Sec. IVB3 below, we show how the summation over all the terms may yield the free-space result.

\subsubsection{TE contribution}
The $z$-dependence of the TE modes [Eq. (\ref{TE})], goes like $K_0(k_{mn}z)$. It is obvious that for small enough $m$ and $n$ and in the limit $z\ll a,b$ discussed here, we have $k_{mn} z\ll 1$.
In order to make our discussion clearer, let us assume $a=b$, then $k_{mn}z=\pi\frac{z}{a}\sqrt{m^2+n^2}$, and for $m,n\ll m^{\ast}\equiv a/(z\pi)$, we indeed have $k_{mn} z\ll 1$ and can take the approximation
\begin{equation}
K_0(k_{mn} z)\approx-\ln(z/ a) -\ln(k_{mn} a)+\ln(2)-\gamma,
\label{TEvdW1}
\end{equation}
where $\gamma\approx0.577$ is Euler's constant.

Let us now try to estimate the total contribution of the TE modes:
as $z/a\rightarrow 0$, the contribution of each $m,n\ll m^{\ast}$ diverges like $\ln(z/a)$. Summing them together, we multiply $\ln(z/a)$ by roughly $m^{\ast}\propto a/z$ terms, ending up with $(1/z)\ln z$ divergence. If we wish to account for any, arbitrary small $z/a$, then $m^{\ast}\rightarrow \infty$ such that our estimation in fact includes the contribution of all terms. To conclude, we estimate
\begin{equation}
\sum_{mn}F^{TE}_{mn,eij}\sim \frac{\ln z}{z}.
\label{TEvdW}
\end{equation}
Using this result in Eq. (\ref{U}), we obtain that the vdW energy mediated by solely TE modes scales as $U\sim (\ln^2 z)/z^2$. Such divergence is much weaker then that of free-space, $1/z^6$, which we expect to obtain when the dipoles are very close. This implies, that the contribution of the TE modes to the dispersion energy at short distances is not the dominant ingredient of the total energy in this regime.

\subsubsection{Comparison to free-space result}
The above conclusion drawn for the TE modes contribution for the dispersion energy suggests that the TM modes are the dominant interaction mediators in the vdW, short-range, regime. We now reside to the calculation of the vdW energy due to TM modes. In order to also relate it to its free-space counterpart, we assume that the dipoles are in the center, namely $x=a/2$ and $y=b/2$, and take $a=b$ for simplicity. Then, for the $i,j=z,z$ component [Eq. (\ref{TMzz})], we obtain,
\begin{eqnarray}
F_{ezz}&=&\sum_{mn}F^{TM}_{mn,ezz}=\frac{4\pi^2}{a^3} \bar{F}_{zz},
\nonumber\\
\bar{F}_{zz}&=&\sum^{\infty}_{m,n=1}\sqrt{m^2+n^2}\sin^2\left(m\frac{\pi}{2}\right)\sin^2\left(n\frac{\pi}{2}\right)
\nonumber\\
&&\times e^{-\sqrt{m^2+n^2}\pi z/a}.
\label{e1}
\end{eqnarray}
We note that $\sin^2(m\pi/2)$ equals $1$ for odd $m$ and $0$ for even $m$, hence the sum $\bar{F}_{zz}$ becomes
$\bar{F}_{zz}=\sum_{m,n=odd}\sqrt{m^2+n^2}e^{-\sqrt{m^2+n^2}\pi z/a}$. For small $z$ we can approximate this sum by an integral, replacing $\sum_{n=odd}\rightarrow (1/2)\int_1^{\infty} dn$ for $n$ and $m$, and obtain
\begin{equation}
\frac{a^3}{4\pi^2}F_{ezz}=\bar{F}_{zz}\approx \frac{1}{4\pi^2(z/a)^3}.
\label{I}
\end{equation}
The integral approximation is tested by comparing it to a direct summation yielding excellent agreement, as can be seen in Fig. 4.

\begin{figure}
\begin{center}
\includegraphics[scale=0.3]{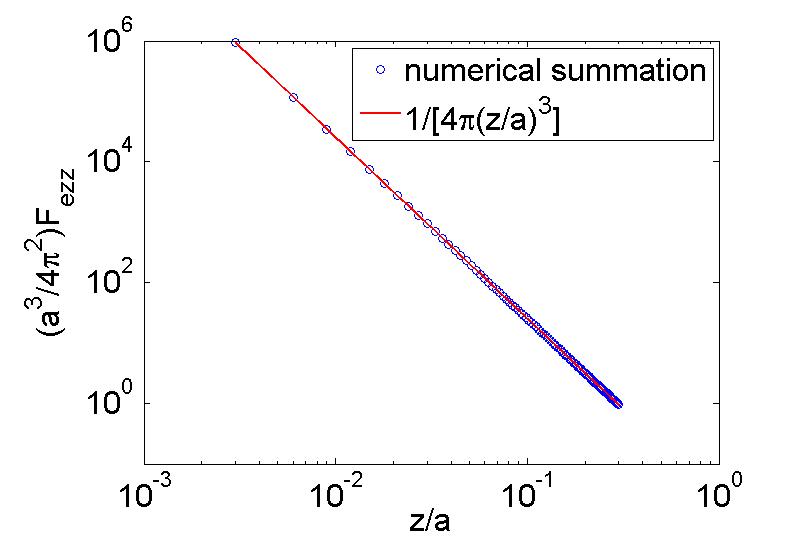}
\caption{\small{(Color online) Evaluation of the sum in Eq. (\ref{e1}): log-log plot of the result of the integral approximation, Eq. (\ref{I}), compared to direct numerical summation. Excellent agreement is observed.
}}
\end{center}
\end{figure}

In a similar fashion we calculate all other cases from Eqs. (\ref{TMzz}),(\ref{TMxx}) and (\ref{TM}) and get
\begin{eqnarray}
F_{ezz}&\approx&\frac{1}{z^3},
\nonumber\\
F_{exx}&=&F_{eyy}\approx-\frac{1}{2z^3},
\nonumber \\
F_{exy}&=&F_{exz}=F_{eyz}=0.
\label{e2}
\end{eqnarray}
Inserting these results into Eq. (\ref{U}), we obtain the dispersion energy in the vdW regime,
\begin{eqnarray}
&&U\approx-\frac{1}{4\pi^2\epsilon^2}\frac{1}{z^6}\sum_{e_1,e_2}\sum_{ijlq} \frac{d_{e_2 i}d_{e_1 j}d_{e_2 l}d_{e_1 q}}{E_{e_1}+E_{e_2}}\times
\nonumber \\
&&  \left[\delta_{ij,zz}-\frac{1}{2}\left(\delta_{ij,xx}+\delta_{ij,yy}\right)\right]
\left[\delta_{lq,zz}-\frac{1}{2}\left(\delta_{lq,xx}+\delta_{lq,yy}\right)\right].
\nonumber\\
\label{UvdW}
\end{eqnarray}
This has to be compared to the vdW result in free-space before averaging over dipole orientations, namely \cite{MQED},
\begin{eqnarray}
U_{fs}&\approx&-\frac{1}{4\pi^2\epsilon^2}\frac{1}{z^6}\sum_{e_1,e_2}\sum_{ijlq} \frac{d_{e_2 i}d_{e_1 j}d_{e_2 l}d_{e_1 q}}{E_{e_1}+E_{e_2}}
\nonumber \\
&&  \times\frac{1}{4}\left[\delta_{ij}-3\hat{R}_i\hat{R}_j\right]\left[\delta_{lq}-3\hat{R}_l\hat{R}_q\right],
\label{fsvdW}
\end{eqnarray}
where $\hat{R}_i=\mathbf{e}_i\cdot\mathbf{e}_z$, i.e. $\hat{R}_z=1$ and $\hat{R}_x=\hat{R}_y=0$. It is easy to verify that Eqs. (\ref{UvdW}) and (\ref{fsvdW}) yield identical results: for $i,j,l,q=z$ the terms in the second row of both equations give 1, for $i,j,l,q=x$ they give 1/4, for $i,j=x$ and $l,q=z$ they give -1/2 (repulsion) and for $i\neq j$ or $l\neq q$ they become $0$.

We conclude that in the vdW regime, namely $z\ll a$, where no wave effects are apparent, and when the dipoles are in the center of the waveguide, the TM modes are dominant and the dispersion energy is that of free-space, as can be expected.

\section{conclusions}
In this paper, we applied ideas drawn from the modification of resonant dipole-dipole interaction in confined geometries, to the study of dispersion interactions between a pair of point-like dipoles. We considered a tightly confined waveguide structure whose transverse-modes cutoff-frequencies are all above the typical frequency of the interacting dipoles. We obtained analytical expressions for the dispersion energy at all inter-dipolar distances $z$. We found that the difference between retarded, wave-like (Casimir-Polder), behavior and  quasistatic (vdW) one, is set  by the confinement length-scale $a$. For $z\ll a$ the interaction is mediated by TM modes and can become identical to its free-space counterpart, whereas in the retarded regime, $z\gg a$, the interaction is exponentially decaying with distance and is carried by $TE_{01}$ and $TE_{10}$ evanescent fields. Hence, the resulting interaction is much shorter range than that in free space. This has an implication on the non-additivity of the vdW and dipole-dipole interactions: in free-space the interaction energy between multiple dipoles is additive, namely, can be obtained by pairwise summation, as long as $\alpha/r^3$ is small, where $\alpha$ and $r$ are typical polarizability and inter-dipolar distance respectively \cite{MIL}. Here however, this $1/r^3$ scaling is expected to change such that even for higher densities, $r<\alpha^{1/3}$, the interaction may still be additive. This entails modifications of, e.g. the effective dielectric constant of a gas of such dipoles, and all Casimir-associated phenomena in a MWG environment, which are based on the retardation and non-additivity \cite{REV}, that are both being altered here.

Finally, let us address the generality of our results. Although they were explicitly derived for a rectangular hollow MWG, the same dependence on distance $z$ is expected for dispersion interactions mediated by any transverse modes with cutoff frequencies higher than that of the dipole transition. These may include, e.g. other cylindrical hollow MWG in various transverse shapes, tightly confined hollow-core fibers and high-order modes of electric transmission lines.

\acknowledgements
We appreciate useful discussions with Grzegorz {\L}ach and Steven Johnson. The support of ISF and DIP is acknowledged.

\end{document}